\documentclass[journal,final,twocolumn,10pt,twoside]{IEEEtran}

\normalsize

\usepackage{graphicx}
\usepackage{amsmath}
\usepackage{booktabs} 
\usepackage{amssymb}
\usepackage{multicol}
\usepackage{epstopdf}
\usepackage[noadjust]{cite}
\usepackage[caption=false,font=footnotesize]{subfig}
\usepackage{relsize}
\usepackage{epstopdf}
\usepackage{makecell}
\usepackage{array}
\usepackage{color}
\usepackage{algorithm}
\usepackage{algcompatible}
\usepackage{balance}

\newcommand{\argmaxA}{\mathop{\mathrm{arg\,max}}} 
\newcommand{\argminA}{\mathop{\mathrm{arg\,min}}} 
\newcommand\norm[1]{\left\lVert#1\right\rVert}

\ifCLASSINFOpdf
\else
\fi

\begin{document}

\title{Index Modulation for Molecular Communication via Diffusion Systems}

\author{Mustafa~Can~Gursoy,~\IEEEmembership{Student~Member,~IEEE,}
	Ertugrul~Basar,~\IEEEmembership{Senior~Member,~IEEE,}
	Ali~Emre~Pusane,~\IEEEmembership{Member,~IEEE,}
	and~Tuna~Tugcu,~\IEEEmembership{Member,~IEEE} 
	\thanks{M. C. Gursoy and A. E. Pusane are with the Department
		of Electrical and Electronics Engineering, Bogazici University, Istanbul, Turkey
		(e-mail: can.gursoy@boun.edu.tr and ali.pusane@boun.edu.tr).}
	\thanks{E. Basar is with the Communications Research and Innovation Laboratory (CoreLab), Department of Electrical and Electronics Engineering, Ko\c{c} University, Sariyer 34450, Istanbul, Turkey (e-mail: ebasar@ku.edu.tr).}
	\thanks{T. Tugcu is with the Department of Computer Engineering, NETLAB, Bogazici University, Istanbul, Turkey (e-mail: tugcu@boun.edu.tr).}}


\markboth{To Appear In IEEE Transactions on Communications}%
{To Appear In IEEE Transactions on Communications}


\maketitle

\begin{abstract}
Molecular communication via diffusion (MCvD) is a molecular communication method that utilizes the free diffusion of carrier molecules to transfer information at the nano-scale. Due to the random propagation of carrier molecules, inter-symbol interference (ISI) is a major issue in an MCvD system. Alongside ISI, inter-link interference (ILI) is also an issue that increases the total interference for MCvD-based multiple-input-multiple-output (MIMO) approaches. Inspired by the antenna index modulation (IM) concept in traditional communication systems, this paper introduces novel IM-based transmission schemes for MCvD systems. In the paper, molecular space shift keying (MSSK) is proposed as a novel modulation for molecular MIMO systems, and it is found that this method combats ISI and ILI considerably better than existing MIMO approaches. For nano-machines that have access to two different molecules, the direct extension of MSSK, quadrature molecular space shift keying (QMSSK) is also proposed. QMSSK is found to combat ISI considerably well whilst not performing well against ILI-caused errors. In order to combat ILI more effectively, another dual-molecule-based novel modulation scheme called the molecular spatial modulation (MSM) is proposed. Combined with the Gray mapping imposed on the antenna indices, MSM is observed to yield reliable error rates for molecular MIMO systems. \\
\end{abstract}


\IEEEpeerreviewmaketitle

\section{Introduction}

\par Molecular communication via diffusion (MCvD) is a bio-inspired molecular communication method that utilizes the diffusive nature of the molecules in fluid environments to convey information among nano-machines \cite{kitap}. In an MCvD system, the information is encoded in the quantity \cite{D-MoSK}, type \cite{CSKMOSK}, temporal position \cite{PPMoriginal}, and possibly more physical properties of the molecular waves. After their release, the messenger molecules diffuse through the channel according to the laws of Brownian motion, and are measured at the receiver end for detection \cite{kitap}. Due to the random propagation of the transmitted molecules, MCvD channels are subject to heavy inter-symbol interference (ISI), which hinders their communication performance \cite{ISI_citeicin}. 

\par Similar to traditional wireless communication systems, multiple-input-multiple-output (MIMO) approaches are also considered in the molecular communications realm with main motivations of increasing system throughput and reducing the bit error rate (BER), at the cost of increased device complexity \cite{molecularMIMOrealized}. One such work introduces repetition coding (RC) and proposes an Alamouti-like coding scheme for a $2\times2$ MIMO MCvD system, and finds that RC yields a desirable diversity gain for such a system, also showing that MIMO approaches indeed provide BER reduction in molecular communications  \cite{molecularMIMOBirkan}. For the receiver end of the considered molecular MIMO link, detection algorithms discussed in \cite{MIMOdetection} are used and comparatively analyzed. As another approach, \cite{MIMOmultiplex} proposes using the multiple available antennas for spatial multiplexing to increase the communication throughput. A macro-scale molecular MIMO system testbed is built and introduced in \cite{molecularMIMOrealized} and \cite{molecularMIMOrealizedDEMO}, experimentally confirming the previous theoretical advantages of introducing MIMO to molecular communications. In addition, \cite{MIMOrealizeNANOTECH} and \cite{MIMOrealizeNANOBIO} realize an Alexa Fluor dye-based lab implementation, introducing another physical testbed for a molecular MIMO scheme.

\par Inspired by its prospects and the opportunities for traditional communications, this paper introduces the IM approach (\cite{EBasarR2,EBasarR3}) to molecular communications as a method to further enhance performance of molecular MIMO systems. Overall, the contributions of the paper are as follows:
\begin{itemize}
	\item Unlike providing diversity or spatial multiplexing with the available antennas as discussed in \cite{molecularMIMOBirkan} and \cite{MIMOmultiplex}, respectively, we propose novel molecular MIMO modulation schemes that use the transmitter antenna indices to encode information bits. 
	\item For the simpler nano-machines that have access to only a single type of messenger molecules, we propose a scheme that utilizes the antenna indices as the only information source. This scheme is referred to as molecular space shift keying (MSSK), due to its resemblance to the space shift keying (SSK) modulation in traditional communication systems \cite{SSK}.
	\item By deriving the theoretical bit error rate expression and through Monte Carlo simulations, we find that MSSK brings great benefits for a molecular MIMO system and provides reliable error performances, as it combats ISI and inter-link interference (ILI) more effectively than the existing molecular MIMO approaches. We also demonstrate the existence of a trade-off between ISI and ILI combating for the proposed IM-based molecular MIMO scheme.
	\item For the systems that have access to two types of molecules, we propose the quadrature molecular space shift keying (QMSSK) scheme as a direct extension of MSSK, similar to the quadrature spatial modulation (QSM) approach presented in \cite{QSMpaper}. 
	\item In order to combat ILI better, we propose another dual-molecule IM-based scheme named the molecular spatial modulation (MSM), a scheme which combines the well-known molecule shift keying (MoSK) scheme with the proposed MSSK. We find that MSM combats ILI-caused errors more effectively than both MSSK and QMSSK, but is subject to more ISI compared to QMSSK.
\end{itemize}

\par One big advantage of the proposed approaches is the fact that only a single antenna (or possibly two for QMSSK) is utilized at a time. Similar to traditional RF-based communications, utilizing fewer antennas for each transmission allows the transmitter to increase the transmission power per channel use, which helps to decrease the relative arrival variance of the messenger molecules, thus the BER. Furthermore, utilizing a single antenna for each transmission eliminates possible synchronization problems among transmit antennas, which may pose a problem in other diversity schemes \cite{EBasarR2}-\cite{EBasarR3}. In terms of computational complexity, the simplicity of the proposed schemes is also more suitable for nano-scale machinery than other diversity schemes. The simplicity of the proposed IM-based schemes is especially prominent in the receiver design, as all of the considered methods are found to yield promising error performances with the maximum count decoder (MCD) considered in the paper, which can be realized using a simple comparator circuit and without the channel impulse response information.

\section{System Model}
\label{systemmodel}

\subsection{General System Topology}
\label{subsec:topology}

\par
\begin{figure}[!t]
	\centering
	\includegraphics[width=0.45\textwidth]{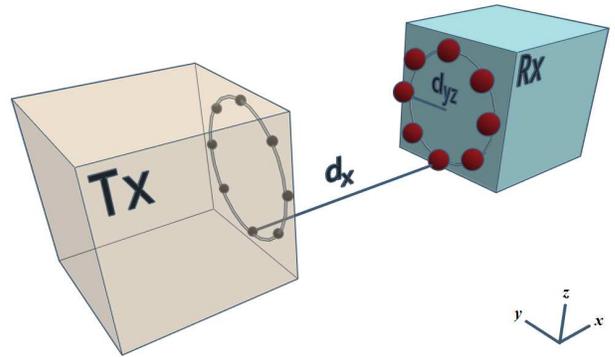}
	\caption{The molecular MIMO system of interest for $n_{Tx} = n_{Rx} = 8$. Each spherical receiver antenna's closest point is $d_{yz}$ away from the center of the UCA, and the receiver antennas of radius $r_r$ are angular-wise $\frac{\pi}{4}$ radians apart from each other. Note that the radius of the transmitter UCA is equal to $d_{yz} + r_r$ for this topology. $d_x$ denotes the closest point of a receiver antenna to its corresponding transmit antenna, and is also equivalent to $d_{Rx-Tx} - 2r_r$ given $d_{Rx-Tx}$ is the distance between the Tx and Rx blocks' surfaces.}	
	\label{fig:UCA}
\end{figure}

 Similar to the system models considered in \cite{molecularMIMOrealized} and \cite{molecularMIMOBirkan}, the system considered in this paper involves a single transmitter block and a single receiver block in an unbounded 3-D MCvD channel environment without drift, as presented in Fig. \ref{fig:UCA}. On the transmitter block's surface (left-hand side of Fig. \ref{fig:UCA}), there are $n_{Tx}$ distinct point sources that work as transmit antennas and are able to emit molecules into the communication channel. When transferring information towards the receiver, the transmitter block unit is assumed to perfectly control the molecule emission of the transmitter antennas, according to the modulation scheme employed. 

\par On the receiver unit's (block's) surface (right-hand side of Fig. \ref{fig:UCA}), there are $n_{Rx}$ spherical absorbing receivers with radii $r_r$, which act as different receiver antennas. In a communication scenario, the receiver block is assumed to collect the number of arrived molecules for each antenna, and perform its decision according to the modulation scheme employed. One thing to note is that the centers of the receiver's spherical antennas are assumed to be perfectly aligned to the corresponding transmitter antennas on the transmitter block. In the paper, the radius of each spherical receiver antenna is chosen to be $r_r = 5 \mu m$.

\par For the scenario considered in this paper, both $n_{Tx}$ and $n_{Rx}$ are chosen as $n_{Tx} = n_{Rx} = 8$, as also shown in Fig. \ref{fig:UCA}. Note that the antennas on both sides are angular-wise equally separated from the center of their respective nano-machines, forming a uniform circular array (UCA) of antennas \cite{UCAreview}. 

The closest distance between the receiver antenna's projection on its surface and the center of the UCA is denoted as $d_{yz}$, which makes the distance between the center of the transmitter antenna and the center of the UCA to be equal to $d_{yz} + r_r$. The closest distance between a transmitter antenna point and its corresponding receiver antenna is denoted by $d_x$. Similar to the topologies considered in \cite{molecularMIMOrealized} and \cite{molecularMIMOBirkan}, the transmitter body is assumed to be fully permeable to the messenger molecules after transmission, whilst the receiver body is assumed to be perfectly reflective, making the molecules elastically collide with its surface if they hit. 

\subsection{The MCvD Channel and the Channel Coefficients}
\label{subsec:channelcoeff}

\par In a 3-D MCvD system without drift, messenger molecules move according to the rules of Brownian motion after their release from the transmitter \cite{kitap}. Using the Fick's diffusion laws, \cite{3Dchar} finds the analytical expression of the molecule arrival distribution with respect to time, for the case of a single point transmitter-single spherical absorbing receiver. Furthermore, \cite{stocasticMult} extends the analysis in \cite{3Dchar} to multiple point transmitters and a single spherical absorbing receiver, and analytically finds the arrival distribution using stochastic geometry. 

\par In the scenario of interest for this paper, the molecular MIMO system at hand consists of multiple transmitters, multiple absorbing receivers, and a reflective surface, as presented in Subsection \ref{subsec:topology}. In the presence of multiple absorbing receivers, extending the work of \cite{3Dchar} directly to multiple antennas results in incorrect modeling of the channel due to the statistical dependence among the arrivals at different receiver antennas. Hence, the channel coefficients of a molecular MIMO system need to be obtained by performing Brownian motion-based Monte Carlo simulations that consider the arrival dependence of the antennas \cite{molecularMIMOrealized}, or by using machine learning methods as mentioned in \cite{molecularMIMOBirkan} and \cite{MLcoeffMIMO}. Afterwards, the arrival to each antenna can be represented by an independent Binomial event with its success probability coming from the appropriate channel coefficient, which was obtained considering the arrival dependence. Hence, to characterize the molecular MIMO channel, the paper firstly uses random-walk-based Monte Carlo simulations to generate the channel response and coefficients for the system of interest.

\par When simulating the messenger molecule propagation with Monte Carlo simulations, time is divided into discrete steps of $\Delta t$, and the position of each molecule in the channel is updated by 
\begin{equation}\label{eq:step}
\begin{aligned}
x(t + \Delta t )&= x(t) + \Delta X, \\
y(t + \Delta t) &= y(t) + \Delta Y, \\
z(t + \Delta t) &= z(t) + \Delta Z \\
\end{aligned}
\end{equation}
for each axis, until it arrives at the receiver and gets absorbed. Here, $\Delta X$, $\Delta Y$, and $\Delta Z$ denote the random incremental steps a molecule takes for each discrete time step in the corresponding axes, and are modeled by the normal random variable $\mathcal{N}(0,2 D \Delta t)$ with mean $0$ and variance $2 D \Delta t$ \cite{kitap}. Also, note that $D$ represents the diffusion coefficient of the messenger molecule and is chosen to be $D = 79.4 \frac{\mu m^{2}}{s}$ throughout the paper, which is considered as a benchmark value in the literature. For sufficient accuracy, the Monte Carlo simulations are performed with $10^6$ molecules and with a time step of $\Delta t = 10^{-4}$ seconds.

\par The time arrival distribution, $f_{hit}(t)$, obtained as a result of the Monte Carlo simulation can be integrated with respect to time to yield $F_{hit}(t)$, the probability of a single molecule's arrival until time $t$. For consequent bit transmissions with a symbol duration of $t_s$, the channel coefficients for a SISO scenario can be found by 
\begin{equation}\label{eq:SISOtaps}
h[n]= F_{hit}\big(nt_s\big) - F_{hit}\big((n-1)t_s\big).
\end{equation}
Note that the transmitter and the receiver are assumed to be synchronized similar to a manner presented in \cite{synchronization}.

\par As the findings of \cite{3Dchar} also suggest, the 3-D MCvD channel's response is heavy tailed and infinite. That is to say, when a molecule is released to the unbounded 3-D communication environment, there exists a non-zero probability that the molecule may never arrive at the receiver end. Hence, the channel memory is infinite, stating the need to have infinitely many $h[n]$'s to perfectly model the channel. However, for all practical purposes, the channel can be modeled with an FIR model, by considering only the first $L$ memory elements \cite{BERconverge}. For an accurate representation of the channel, this paper considers channel memory $L=30$. 

\par In consequent bit transmission scenarios, the transmitted molecules may arrive at symbol intervals other than the intended interval, causing ISI for MCvD systems. Furthermore, the MIMO nature of the system of interest in this paper also brings ILI into the system and requires consideration of the channel responses for each transmitter and receiver antenna combination separately. Throughout the paper, the $n^{th}$ channel coefficient of the subchannel corresponding to the $i^{th}$ transmitter and $j^{th}$ receiver is denoted as $h_{i,j}[n]$. As an example, Table \ref{tab:coefs} shows the first five channel coefficients $h_{1,j}[n]$ where $j=1,...,8$ and $n=1,...,5$.

\begin{table}[!h]
	\centering
	\caption{First five channel coefficients on all $n_{Rx} = 8$ receivers when the antenna with index number $1$ transmits. $n_{Tx} = n_{Rx} = 8$, $d_{x} = 10\mu m$, $d_{yz} = 10\mu m$, $r_r = 5 \mu m$, $D = 79.4 \frac{\mu m^{2}}{s}$, and $t_s = 0.75$s.}
	\label{tab:coefs}
	\begin{tabular}{ | c | c | c | c | c | c |}
		\hline
		\makecell{Time (horizontal)\\Space (vertical)} & $h_{1,j}[1]$ & $h_{1,j}[2]$ & $h_{1,j}[3]$ & $h_{1,j}[4]$ & $h_{1,j}[5]$ \\ \hline
		$h_{1,1}[n]$ & $0.1042$ & $0.0346$ & $0.0141$ & $0.0078$ & $0.0049$ \\ \hline
		$h_{1,2}[n]$ & $0.0357$ & $0.0227$ & $0.0106$ & $0.0062$ & $0.0039$ \\ \hline
		$h_{1,3}[n]$ & $0.0052$ & $0.0090$ & $0.0057$ & $0.0036$ & $0.0026$ \\ \hline
		$h_{1,4}[n]$ & $0.0014$ & $0.0045$ & $0.0033$ & $0.0023$ & $0.0017$ \\ \hline
		$h_{1,5}[n]$ & $0.0009$ & $0.0035$ & $0.0029$ & $0.0021$ & $0.0014$ \\ \hline
		$h_{1,6}[n]$ & $0.0014$ & $0.0045$ & $0.0033$ & $0.0023$ & $0.0017$ \\ \hline
		$h_{1,7}[n]$ & $0.0052$ & $0.0090$ & $0.0057$ & $0.0036$ & $0.0026$ \\ \hline
		$h_{1,8}[n]$ & $0.0357$ & $0.0227$ & $0.0106$ & $0.0062$ & $0.0039$ \\
		\hline
	\end{tabular}
\end{table}

\par Note that the channel coefficients presented in Table \ref{tab:coefs} can be interpreted as the channel coefficients when a transmission is made from the antenna with index number $1$. One thing to infer from Table \ref{tab:coefs} is the fact that the receiver antennas that are equidistant to the receiver antenna with index $1$ have the same channel coefficients. The reasons for this lie in the assumption that each transmit antenna is aligned with the center of its corresponding receiver antenna, and the fact that the antennas are placed to form a UCA. Another implication of the UCA antenna deployment is the spatial symmetry it brings to the system. For example, note that the UCA deployment implies $h_{1,j}[n] = h_{2,(j+1)}[n] = h_{3,(j+2)}[n]$, etc. To generalize, it can accurately be stated that the channel coefficients when a transmission is made from the $i^{th}$ transmitter antenna is equivalent to circularly shifting the columns of Table \ref{tab:coefs} by $(i-1)$. This phenomenon brings a useful simplification when simulating the system impulse response presented in this paper: The channel can be modeled correctly by considering only the response of a single transmitter. However, it should be noted at this point that all the channel model equations, analytical derivations, and receiver operations in the paper are expressed in a generalized manner, rather than specifically for the UCA arrangement. The UCA arrangement is used solely for demonstrative purposes in the paper, and the proposed schemes can be used under any other antenna geometry.

\par When transmitting multiple molecules from a transmitter antenna, the true and exact arrival counts at the receiver antennas need to be characterized by a joint distribution among all $n_{Rx}$ receivers due to the statistical dependence between antenna arrivals. However, stemming from the fact that this dependence is accounted for when generating the channel coefficients using the aforementioned random-walk-based Monte Carlo simulations, this paper uses the approach employed in  \cite{molecularMIMOrealized,molecularMIMOBirkan,MIMOdetection} and approximates the arrival counts at each receiver antenna as an independent Binomial random variable with success probability $h_{i,j}[n]$. Furthermore, the channel parameters presented in Table \ref{tab:coefs} let the Gaussian approximation of Binomial arrivals be sufficiently accurate as stated in \cite{arrivalmodel}. Therefore, the Gaussian approximation is usable for the scenarios in this paper. It is also noteworthy that the received number of molecules for each receiver antenna is the sum of all $n_{Tx}$ transmit antenna responses in this paper, as a direct extension to the arguments presented in \cite{molecularMIMOBirkan} and \cite{stocasticMult}. Overall, the total number of molecules at the $j^{th}$ receiver antenna arriving at the $k^{th}$ symbol interval $R_j[k]$ can be approximated by $R_j[k] \sim \mathcal{N}(\mu_j[k],\sigma^2_j[k])$, where
\begin{equation}\label{eq:gausmean}
\mu_j[k] = \sum_{z=k-L+1}^{k} \sum_{i=1}^{n_{Tx}} s_{i}[z] h_{i,j}[k-z+1]
\end{equation}
and 
\begin{equation}\label{eq:gausvar}
\begin{split}
\sigma^2_j&[k] = \\ & \sum_{z=k-L+1}^{k} \sum_{i=1}^{n_{Tx}} s_{i}[z] h_{i,j}[k-z+1] \big(1-h_{i,j}[k-z+1]\big).
\end{split}
\end{equation}
In expressions \eqref{eq:gausmean} and \eqref{eq:gausvar}, $s_{i}[z]$ denotes the modulation mapping of the $z^{th}$ symbol on the $i^{th}$ transmit antenna. It can also be thought of as the transmitted number of molecules from the $i^{th}$ transmit antenna on the given symbol interval.

\section{The SISO Baseline System and Molecular MIMO Approaches}
\label{sec:tobecompared}

\par As discussed in several previous works, including but not limited to \cite{molecularMIMOrealized}, \cite{molecularMIMOBirkan}, and \cite{MIMOmultiplex}, the MIMO concept, which is a vital part of many modern RF-based wireless communication systems, provides promising results in terms of throughput and error performance when applied on the molecular communication realm. This section presents the existing space-time coding and spatial multiplexing methods for molecular MIMO systems, alongside the SISO baseline used for comparison.

\subsection{SISO Baseline}
\label{subsec:SISO}

\par As the SISO baseline scheme, SISO communication using the on-off keying (OOK) version of binary concentration shift keying (BCSK) is used in this paper \cite{CSKMOSK}. BCSK is the quantity modulation equivalent of molecular communications systems, and transmits a bit-1 by transmitting $s[k] = M^{Tx}$ molecules, and a bit-0 by transmitting no molecules ($s[k] = 0$). 

\par The synchronized receiver nano-machine counts the arriving molecules until the end of the symbol duration, and compares the said arrival count $R[k]$ with a threshold $\gamma$ to decode the transmitted bit. This decoder is referred to as the fixed threshold decoder (FTD) throughout the paper. Note that, as can also be recalled from Subsection \ref{subsec:channelcoeff}, the molecules that arrive in later symbol intervals are the main sources of ISI in an MCvD system.

\par It is noteworthy that the data rate and the energy consumption need to be normalized among different schemes. With a normalized bit rate of $\frac{1}{t_b}$, the symbol duration for a SISO BCSK transmission is $t_s = t_b$, since only one bit is transmitted at a time. Furthermore, due to the relation between the energy consumption of a molecular communication scheme and the number of transmitted molecules, the energy consumption per bit constraint is equivalent to a constraint imposed on the transmitted number of molecules per bit \cite{energymodel,BurcuISI}. For a SISO BCSK scheme, $\frac{1}{2}M^{Tx}$ molecules are transmitted on the average, since the probability of transmitting a bit-1 is assumed to be $0.5$ in this paper. Therefore, the constraint of transmitting $\frac{1}{2}M^{Tx}$ molecules per bit on average is imposed on the considered schemes.

\subsection{Repetition Coding}

\par The study of \cite{molecularMIMOBirkan} introduces the RC scheme to the molecular MIMO literature, using the aforementioned BCSK as the employed modulation scheme. The transmission vector of such a scheme is defined as
\begin{equation}\label{eq:repetitionG}
\textbf{g}_{\textnormal{RC}} =
\overbrace{\Big[s[k] \hspace{0.2cm} s[k] \hspace{0.2cm} \cdots \hspace{0.2cm}  s[k]\Big]}^\text{$n_{Tx}$}
\end{equation}
where $s[k]$ denotes the mapping of the $k^{th}$ transmitted BCSK symbol.

\par At the receiver end, selection combining (SC) and equal gain combining (EGC) are considered. Denoting the received number of molecules corresponding to the $j^{th}$ receiver antenna for the $k^{th}$ symbol in the sequence to be transmitted as $R_{j}[k]$, the total number of received molecules for the selection combining method is found by
\begin{equation}\label{eq:selectionrec}
R_{\text{SC}}[k] = \max(R_{1}[k],\cdots,R_{n_{Rx}}[k]).
\end{equation}
Furthermore, the total number of received molecules for the EGC method can be expressed as
\begin{equation}\label{eq:EGCrec}
R_{\text{EGC}}[k] = \sum_{j=1}^{n_{Rx}}R_{j}[k].
\end{equation}
Note that since the UCA nature of the antennas implies symmetric channel coefficients for each receiver, the maximum-ratio combining (MRC) is equivalent to EGC for the system considered in this paper, similar to \cite{molecularMIMOBirkan}. 

\par Even though the data symbol is replicated $n_{Tx}$ times in the space axis, this scheme still transmits a single bit per its unit symbol duration. Hence, the symbol duration for this scheme becomes simply $t_b$. Furthermore, since the same bit is repeated $n_{Tx}$ times in the space axis, the total budget of transmitting $M^{Tx}$ molecules for a bit-1 in SISO BCSK needs to be divided into $n_{Tx}$ equal transmissions for normalization. Consequently, in RC, each antenna transmits $\frac{M^{Tx}}{n_{Tx}}$ molecules for a bit-1 to satisfy the energy consumption constraint. Hence, when employing BCSK for modulation, the number of transmitted molecules, $s[k]$, in \eqref{eq:repetitionG} becomes $\frac{M^{Tx}}{n_{Tx}}$ if the $k^{th}$ bit in the sequence is a bit-1 ($u[k]=1$), and becomes $0$ if $u[k]=0$.

\subsection{Spatial Multiplexing}

\par As initially introduced to the molecular communication literature by \cite{MIMOmultiplex}, spatial multiplexing (SMUX) aims to increase the overall system throughput by dividing the bit sequence to be transmitted into $n_{Tx}$ parallel streams, and transmitting these different streams from $n_{Tx}$ different transmit antennas. Thus for molecular SMUX, the transmission vector  can be expressed as
\begin{equation}\label{eq:SMUX_G}
\textbf{g}_{\textnormal{SMUX}} =
\overbrace{\Big[s[kn_{Tx}+ 1] \hspace{0.2cm} s[kn_{Tx} + 2] \hspace{0.2cm}  ... \hspace{0.2cm}  s[kn_{Tx}+k]\Big]}^\text{$n_{Tx}$},
\end{equation}
similar to SMUX in traditional communication systems.

\par With a molecular MIMO system with $n_{Tx} = n_{Rx}$, each receiver antenna counts the number of molecules it receives, and performs a threshold based detection for each bit. Note that the transmitter antenna with index $i$ is \textit{paired} with the $i^{th}$ receiver antenna for this scheme, so the detection done at the $i^{th}$ receiver antenna estimates the bit transmitted from the $i^{th}$ transmitter antenna. 

\par This scheme transmits $n_{Tx}$ different bits in parallel using each of its transmitter antenna-receiver antenna pair. This allows the SMUX scheme to transmit with a symbol duration of $n_{Tx}\hspace{0.4mm} t_b$, which helps greatly to combat with ISI. Furthermore, since every parallel subchannel use carries a single bit, the SMUX-BCSK scheme can represent a bit-1 with $s[k] = M^{Tx}$ molecules for each transmitter antenna to satisfy the energy constraint.

\section{Proposed Methods}
\label{sec:IMbased}

\par SMUX is able to combat ISI very effectively by increasing the symbol duration while keeping the bit rate constant. However, since each transmitter-receiver antenna pair conveys different pieces of information, this scheme suffers from heavy ILI. Furthermore, methods like RC and SMUX require perfect synchronization between the transmitter antennas, which may be a cumbersome task for a simple nano-machine. Motivated by these potential shortcomings and the benefits of spatial modulation approaches \cite{EBasarR1}-\cite{EBasarR4}, this section introduces novel antenna index-based modulation schemes to the molecular communications realm. 

\subsection{Molecular Space Shift Keying (MSSK)} 
\label{subsec:single}

\par Similar to the SSK scheme as introduced by \cite{SSK}, MSSK uses the antenna index as the \textit{only} way to convey information. In an $n_{Tx}$-MSSK scheme, each antenna represents a $\log_2 n_{Tx}$-bit string. In the scheme, the transmission is done by dividing the original bit sequence $\textbf{u}$ into groups of $\log_2 n_{Tx}$ bits, mapping the $\log_2 n_{Tx}$-bit long string to its appropriate transmit antenna, and \textit{activating} only that antenna for transmission while keeping others idle. Since every channel transmission represents $\log_2 n_{Tx}$ bits (assuming $n_{Tx}$ is an integer power of $2$), the transmitter is able to send $s_{\text{MSSK}} = \frac{\log_2 n_{Tx}}{2}M^{Tx}$ molecules from the activated antenna with a symbol duration of $(\log_2 n_{Tx}) t_b$, while satisfying both the energy consumption and bit rate constraints. For the sake of clear presentation, this paper considers a molecular MIMO system with $n_{Tx} = 8$ transmitter and $n_{Rx} = 8$ receiver antennas, which allows encoding $3$ bits using the antenna index.

\par At the receiver end, the receiver counts the number of arrivals to each antenna until the end of the symbol duration, and decides on the maximum arrival among antennas. Denoting the transmitted symbol (hence the activated antenna) for the $k^{th}$ transmission instant as $x[k]$, the receiver decodes $x[k]$ by performing 
\begin{equation}\label{eq:MSSKmax}
\hat{x}[k] = \argmaxA_{{j} \in \{1,\cdots,n_{Rx}\}}R_{j}[k].
\end{equation}
After the estimation of $x[k]$, the receiver then maps $\hat{x}[k]$ onto the appropriate $\log_2 n_{Tx}$-bit sequence to decode the original bit sequence. This decoding method is referred to as the maximum count detector (MCD) throughout the paper, and is a widely used decoding method for molecular communication systems due to its simplicity and ability to work without the channel impulse response (CIR). Note that, for the system considered in this paper, CIR corresponds to the channel coefficient matrix presented in Table \ref{tab:coefs} \cite{bayramPPM}.

\subsubsection*{Gray Mapping} 

\par
\begin{figure}[h]
	\centering
	\includegraphics[width=0.45\textwidth]{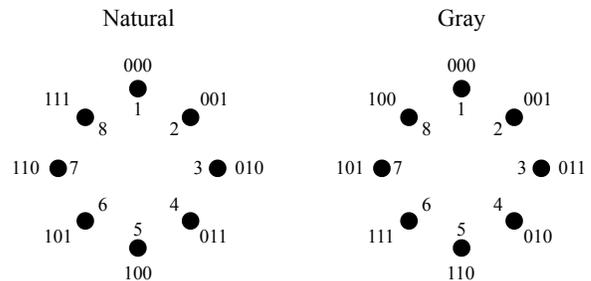} 
	\caption{Antenna indices (inside) and corresponding bit sequences (outside) for natural mapped MSSK (left) and Gray mapped MSSK (right) modulations.}
	\label{fig:binary_vs_gray}
\end{figure}

 Since the receiver performs maximum count detection for MSSK, the possible error sources can be caused by both ISI and ILI. It can be inferred from the vertical axis of Table \ref{tab:coefs} that most prominent ILI-caused errors are due to the two adjacent receiver antennas of the intended one. In order to reduce the number of bit errors due to ILI for MSSK, the antenna indices can be incorporated with Gray coded indices. MSSK's natural binary mapping and gray mapping of the antennas  are shown in Fig. \ref{fig:binary_vs_gray}.

\par Note that since the decoding is done symbol-wise, the MCD works exactly the same for the Gray coded variant's decoder. The only difference of the Gray coded variant's decoder is the extra block that maps the decoded antenna index $\hat{x}[k]$ to the appropriate bit string according to the Gray code.

\subsection{Dual-Molecule Index Modulation Schemes}
\label{subsec:two}

\par When the system has two types of molecules in hand, there are naturally more possibilities modulation-wise, since the second molecule adds another degree of freedom to the system. Firstly and naturally, the discussion made for the single-molecule scenarios can be directly extended by utilizing the two types of molecules as two orthogonal channels. This applies to all SISO baseline, space-time coding, SMUX, and proposed index-based modulation schemes. Note that the SISO baseline for dual-molecule systems is the binary depleted molecular shift keying (D-MoSK) modulation presented in \cite{D-MoSK}, which is defined as two BCSK streams working in parallel and orthogonal channels. Considering the applicability of the second molecule on molecular IM approaches, this subsection introduces the dual-molecule IM-based schemes.

\subsubsection{Quadrature Molecular Space Shift Keying (QMSSK)}

\par QMSSK consists of two parallel MSSK modulators to convey information towards the receiver nano-machine. This method is a direct extension of the $n_{Tx}$-MSSK modulation for two types of molecules, utilizing the fact that two available molecules provide two orthogonal channels for use.

\par In an $n_{Tx}$-QMSSK scheme, the transmitter groups the bit stream into groups of $2\log_2 n_{Tx}$, where first $\log_2 n_{Tx}$ bits are encoded by performing $n_{Tx}$-MSSK with molecule type-A and the last $\log_2 n_{Tx}$ bits are encoded with type-B. Note that since the system can send $2\log_2 n_{Tx}$ bits per transmission, the symbol duration can be doubled to reach $(2\log_2 n_{Tx})t_b$ for this scheme, which helps greatly in terms of ISI combating. Also, the system releases $s_{\text{QMSSK}} = \frac{\log_2 n_{Tx}}{2}M^{Tx}$ molecules per transmission per molecule type, since every channel use per molecule type conveys $\log_2 n_{Tx}$ bits.

\par Similar to MSSK, QMSSK is also compatible with the MCD, where the $\argmaxA$ operation in \eqref{eq:MSSKmax} is performed separately for type-A and type-B molecules to detect the symbols. Furthermore, Gray mapping is also applicable for $n_{Tx}$-QMSSK as well. 

\subsubsection{Molecular Spatial Modulation (MSM)}

\par Transmitting two parallel streams of $n_{Tx}$-MSSK with $n_{Tx}$-QMSSK helps the system greatly by combating ISI with an increased symbol duration of $(2\log_2 n_{Tx})t_b$. However, like MSSK, QMSSK is also prone to ILI-caused errors since a transmission aimed towards a certain receiver antenna also causes molecule arrivals to the adjacent antennas. 

\par Instead of creating two orthogonal and parallel streams as in QMSSK, the two available molecules may be used to perform binary type-modulation (binary MoSK, BMoSK) as well. Combining BMoSK with MSSK yields a new family of index-based modulation schemes, the molecular spatial modulation (MSM), in which the transmitter separates the $\textbf{u}$ sequence in groups of $1 + \log_2 (n_{Tx})$, encodes the first bit using BMoSK, and the remaining $\log_2(n_{Tx})$ bits using $n_{Tx}$-MSSK. 

\par At the receiver end, MCD presented in \eqref{eq:MSSKmax} can be directly extended to incorporate both type-A and type-B molecules. The presented maximum count decoder in \eqref{eq:MSMmax} decodes the activated antenna index $x[k]$ as
\begin{equation}\label{eq:MSMmax}
\begin{split}
\hat{x}[k] = \argmaxA_{{j} \in \{1,\cdots,n_{Rx}\}}  \Bigg( \max \big( R^A_{j}[k],R^B_{j}[k] \big) \Bigg),
\end{split}
\end{equation}
where $R^A_{j}[k]$ and $R^B_{j}[k]$ denote the received number of type-A and type-B molecules to the $j^{th}$ receiver antenna, respectively. Similar to $\hat{x}[k]$, the binary MoSK-encoded single bit is also decoded with a maximum count operation \cite{isomermosk} for this MCD. The decoded bit can be combined with the symbol mapping of $\hat{x}[k]$ to decode the full transmitted sequence of $1 + \log_2 (n_{Tx})$ bits.

\par Since it only transmits a single type of molecule per transmission, the introduction of BMoSK to MSM serves to reduce ILI by helping the channel \textit{clean} itself from the other type of molecule when it is not transmitted. However, utilizing the second molecule to encode only a single bit instead of $\log_2 n_{Tx}$ as in QMSSK makes MSM suffer from higher ISI since it is allowed to transmit at a symbol duration of $(\log_2 (n_{Tx}) + 1)t_b$. Note that this is valid when $(\log_2 (n_{Tx}) + 1) < 2\log_2 (n_{Tx})$, which holds for the system presented in this paper, since $n_{Tx} = n_{Rx} = 8$. Hence, what MSM provides can be considered as better ILI combating at the cost of worse ISI combating. 

\par In the presence of more molecules, the MSM approach can be extended directly. Additional molecule types can help MSM to combat ILI even more with increased orders of MoSK, but they come with a cost of increased complexity in nano-machine circuitry. Overall, an MSM scheme with $\beta$ different molecules and $n_{Tx}$ antenna indices can be denoted as ($\beta , n_{Tx}$)-MSM. In this paper's scenario of interest, the considered MSM scheme is chosen as ($2,8$)-MSM for demonstrative purposes. Gray mapping for antenna indexing is also applicable for MSM.

\section{Theoretical Bit Error Rate}
\label{sec:theo}

\par Since an MCvD system is subject to signal-dependent noise and ISI, theoretical BER expressions for MCvD systems require evaluation over all possible symbol sequences with length $L$ \cite{arrivalmodel}. MSSK is no exception to this, and the theoretical BER $P_e$ can be found by 
\begin{equation}\label{eq:overall}
P_e = \sum_{\forall x[k-L+1:k]}^{} \Big(\frac{1}{n_{Tx}}\Big)^L  P_{e|x[k-L+1:k]}
\end{equation}
where $x[k-L+1:k]$ denotes the activated antenna index sequence between $(k-L+1)^{th}$ and $k^{th}$ transmission, both inclusive. Each element of $x[k-L+1:k]$ is an element of the set $\{1,2,\cdots,n_{Tx}\}$. Furthermore, $P_{e|x[k-L+1:k]}$ represents the probability of error when the sequence of activated antennas is $x[k-L+1:k]$, and can be expressed as
\begin{equation}\label{eq:conditional}
P_{e|x[k-L+1:k]} = \sum_{j=1}^{n_{Rx}} P(R_j[k] > R_{j}^{'}[k]) \frac{d_H(\textbf{v}_{\textbf{x[k]}},\textbf{v}_{\textbf{j}})}{\log_2(n_{Tx}) }  
\end{equation}
where $R_{j}^{'}[k]$ represents the arrival counts corresponding to all receiver antennas other than the $j^{th}$. In addition, $d_H(\cdot)$ represents the Hamming distance operator that finds the number of differing bits between two bit sequences. In this case of interest, these sequences are $\textbf{v}_{\textbf{x[k]}}$ and $\textbf{v}_{\textbf{j}}$, the $\log_2 (n_{Tx})$-bit codeword vectors that correspond to the antenna indices $x[k]$ and $j$, respectively, given $n_{Tx}$ is an integer power of $2$. 

\par The expression $P(R_j[k] > R_{j}^{'}[k])$ in \eqref{eq:conditional} denotes the probability that the molecule arrivals to the $j^{th}$ antenna is the largest among all arrivals to other receiver antennas at the $k^{th}$ symbol interval. Hence, the right-hand side of \eqref{eq:conditional} computes the probability of a certain antenna receiving the maximum number of molecules, and multiplies that probability with the bit error rate given that antenna is chosen by the MCD. Overall, this weighted sum can be interpreted as an expectation over the probability mass function of \textit{$j^{th}$ antenna receiving the most molecules}, and yields the expected bit error rate conditioned on a certain antenna transmission sequence $x[k-L+1:k]$. Also, note that \eqref{eq:conditional} is a generalized expression and can be applied to both natural and Gray mapped MSSK schemes.

\par Recalling from Subsection \ref{subsec:channelcoeff} that the arrival count to each antenna is approximated by a normally distributed random variable, whose mean and variance come from \eqref{eq:gausmean} and \eqref{eq:gausvar}, respectively, $P(R_j[k] > R_{j}^{'}[k])$ is the probability of a normally distributed random variable being greater than all other ($n_{Tx}-1$) normally distributed random variables. Thus, the probability $P(R_j[k] > R_{j}^{'}[k])$ can be re-written as
\begin{equation}\label{eq:maxli}
\begin{split}
P(R_j[k] > R_{j}^{'}[k]) = P\Big(\max &(R_{\tau}[k]) < R_j[k] \Big), \\
 &\forall \tau \in \{1,\cdots,n_{Rx}\} \backslash \{j\}.
\end{split}
\end{equation}
The probability $P(R_j[k] > R_{j}^{'}[k])$ can be obtained by averaging the probability of all $R_{\tau}[k]$'s being smaller than a dummy variable $r$, which obeys the probability density function (PDF) of $R_j[k]$. Hence, $P(R_j[k] > R_{j}^{'}[k])$ can be found by
\begin{equation}\label{eq:approx1}
P(R_j[k] > R_{j}^{'}[k]) = \int_{-\infty}^{\infty} \Bigg[ \prod_{\substack{\tau = 1 \\ \tau \neq j}}^{n_{Rx}} P\big(R_{\tau}[k] < r \big) \Bigg] f_{R_j[k]}(r) dr
\end{equation}
similar to the approach presented in \cite{MaaFjournal}. $f_{R_j[k]}(r)$ denotes the PDF corresponding to $R_j[k]$. Since all antenna arrivals are approximated to be normally distributed, $P\big(R_{\tau}[k] < r \big)$ can be related to the tail distribution of the standard normal distribution (the $Q$-function) as $1-Q\Big(\frac{r-\mu_{\tau}[k]}{\sqrt{\sigma_{\tau}^2[k]}}\Big)$ and $f_{R_j[k]}(r)$ is the normal PDF with the appropriate mean and variance. Overall, $P(R_j[k] > R_{j}^{'}[k])$ can be written as 
\begin{equation}\label{eq:approx2}
\begin{split}
P(R_j[k] > R_{j}&^{'}[k]) =\\ &\int_{-\infty}^{\infty} \Bigg[ \prod_{\substack{\tau = 1 \\ \tau \neq j}}^{n_{Rx}} 1-Q\Big(\frac{r-\mu_{\tau}[k]}{\sqrt{\sigma_{\tau}^2[k]}}\Big) \Bigg] f_{R_j[k]}(r) dr
\end{split}
\end{equation}
where $f_{R_j[k]}(r) = \frac{1}{\sqrt{2\pi \sigma^2_j[k]}} e^{\frac{-(r-\mu_j[k])^2}{2\sigma^2_j[k]}}$, the corresponding normal PDF. By first plugging \eqref{eq:approx2} into \eqref{eq:conditional}, then \eqref{eq:conditional} into \eqref{eq:overall}, the theoretical error probability of the $n_{Tx}$-MSSK scheme under the Gaussian arrival approximation, when $n_{Tx} = n_{Rx}$ can be expressed as
\begin{equation}\label{eq:endTheo}
\begin{split}
P_e = & \Big(\frac{1}{n_{Tx}}\Big)^L \hspace{-0.4cm} \mathlarger{\mathlarger{\sum}}_{\forall x[k-L+1:k]} \mathlarger{\sum}_{j=1}^{n_{Rx}} \Bigg( \frac{d_H(\textbf{v}_{\textbf{x[k]}},\textbf{v}_{\textbf{j}})}{\log_2(n_{Tx}) } \\ &\int_{-\infty}^{\infty} \bigg[ \prod_{\substack{\tau = 1 \\ \tau \neq j}}^{n_{Rx}} Q\Big(\frac{\mu_{\tau}[k]-r}{\sqrt{\sigma_{\tau}^2[k]}}\Big) \bigg] \frac{1}{\sqrt{2\pi \sigma^2_j[k]}} e^{\frac{-(r-\mu_j[k])^2}{2\sigma^2_j[k]}}dr  \Bigg) .
\end{split}
\end{equation}

\par
\begin{figure}[h]
	\centering
	\includegraphics[width=0.48\textwidth]{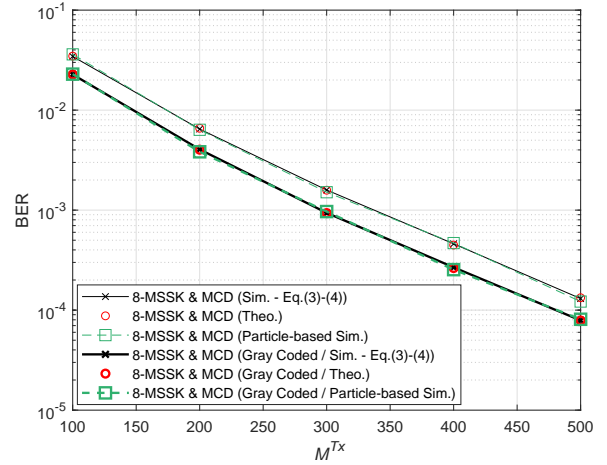} 
	\caption{Simulation-based and theoretical BER vs. $M^{Tx}$ curves of $8$-MSSK for both natural and Gray mapping. $t_b =0.25$s, $d_x = 10\mu$m, $d_{yz} = 10\mu$m, $D = 79.4 \frac {\mu m^{2}}{s}$, $r_r = 5\mu$m, and $L=5$.}
	\label{fig:compBER}
\end{figure}

 Finding the theoretical BER with \eqref{eq:endTheo} requires the computation of all possible $x[k-L+1:k]$ sequences. Since there are $(n_{Tx})^{L}$ different combinations, this is an extremely computationally complex task, requiring $8^{30} \approx 1.2 \times 10^{27}$ evaluations of \eqref{eq:conditional} for the scenario of interest in this paper. Under the light of this finding, a comparative analysis of the BER curves obtained by computer simulations and evaluating \eqref{eq:endTheo} for a shorter channel memory of $L=5$ is presented in Fig. \ref{fig:compBER}, for demonstrative purposes. In Fig. \ref{fig:compBER}, the theoretical BER obtained by \eqref{eq:endTheo} is comparatively analyzed with both particle-based simulations described by Algorithm \ref{alg:particle} in the Appendix (similar to \cite{particlebased} and \cite{particle-birkan}) and simulations made on the channel model presented in Section \ref{systemmodel}. For both of the simulation methods, the same channel topology with equivalent channel, system, and communication parameters is considered, including the channel memory $L=5$.

\par Note that the exact same analysis holds for $n_{Tx}$-QMSSK as well, and a similar analysis can directly be extended for the ($\beta,n_{Tx}$)-MSM scheme. Since there are $\beta$ molecule types and $n_{Tx}$ transmit antennas for a ($\beta,n_{Tx}$)-MSM scheme, the same analysis needs to be done by considering all $(\beta \hspace{0.1cm} n_{Tx})^{L}$ cases instead of the $(n_{Tx})^{L}$ for $n_{Tx}$-MSSK. It is also noteworthy that \eqref{eq:endTheo} is a general theoretical BER expression for MSSK that is applicable to all antenna geometries, rather than a specific one for the scenario considered in this paper.

\section{Error Performance Evaluation}
\label{sec:errorperformance}

\par We comparatively analyze the BER performances of the proposed systems with the help of computer simulations using the channel model described in Subsection \ref{subsec:channelcoeff}, alongside the existing molecular MIMO methods in this section. Since using multiple molecules is considered as a complexity burden for molecular communication systems, the methods for single and two types of molecules are analyzed separately for fairness.

\par In the performed computer simulations, the default values of system and channel parameters are chosen as in Table \ref{tab:exppar1}. If a parameter is not the swept simulation parameter, its value is equal to the value presented in Table \ref{tab:exppar1}. This is valid for both single and dual-molecule scenarios.
\begin{table}[h]
	\centering
	\caption{Default System and Channel Parameters for Single Molecule Scenarios.}
	\label{tab:exppar1}
	\begin{tabular}{@{}llll@{}}
		\toprule
		Parameter Symbol   & Default Value    &  &  \\ \midrule
		$r_r$  &   $5\mu m$                    &  &  \\
		$d_x$  &   $10\mu m$                   &  &  \\
		$d_{yz}$ & $10\mu m$                   &  &  \\
		$D$    &  $79.4 \frac {\mu m^{2}}{s}$ &  &  \\ 
		$L$	& $30$ symbols &  &  \\
		$M^{Tx}$ & $300$ molecules & & \\
		$t_b$ & $0.25$s & & \\ \bottomrule
	\end{tabular}
\end{table}

\subsection{Single-Molecule Systems}
\label{subsec:errorsingle}

\par Recalling that the proposed single-molecule IM-based scheme is referred to as MSSK in Subsection \ref{subsec:single}, this subsection aims to comparatively analyze MSSK's BER performance with other molecular MIMO schemes, under different channel and system conditions. To compare the BER performance of MSSK, the SMUX and RC schemes as presented in Section \ref{sec:tobecompared}, are considered. Note that since there are $n_{Tx} = n_{Rx} = 8$ transmitter and receiver antennas, the SMUX scheme creates $8$ parallel streams, RC repeats the symbols at all $8$ transmitter antennas, and $8$-MSSK is employed as the IM-based approach.

\par
\begin{figure}[h]
	\centering
	\includegraphics[width=0.48\textwidth]{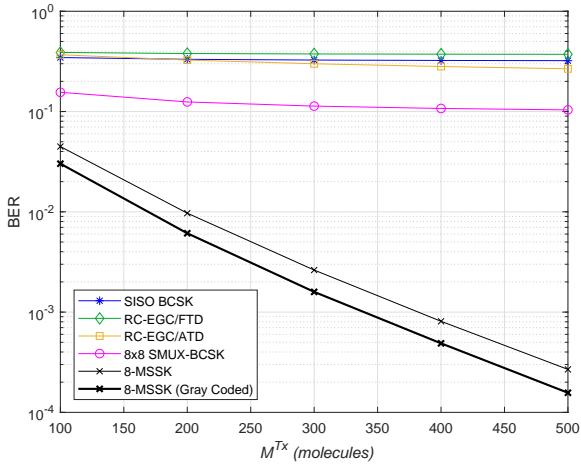} 
	\caption{BER vs. $M^{Tx}$ curves for the single-molecule MIMO approaches. $t_b =0.25$s, $d_x = 10\mu$m, $d_{yz} = 10\mu$m, $D = 79.4 \frac {\mu m^{2}}{s}$, and $r_r = 5\mu$m. }
	\label{fig:single_Msweep}
\end{figure}

 At the receiver end of the RC scheme, FTD presented in Subsection \ref{subsec:SISO} and the adaptive threshold decoder (ATD) as mentioned in \cite{ATDpaper} and \cite{molecularMIMOBirkan} are employed. For SMUX, the receiver is assumed to perform fixed threshold decoding, as described in \cite{CSKMOSK} to decode BCSK modulated symbols. Overall, Fig. \ref{fig:single_Msweep} shows the BER vs. $M^{Tx}$ curve for the considered schemes in a system with parameters as in Table \ref{tab:exppar1}.

\par Firstly, Fig. \ref{fig:single_Msweep} shows that the RC combined with FTD and ATD yields similar BER results compared to the SISO baseline. Additionally, ATD performs better than FTD and surpasses the SISO baseline, which agrees with \cite{molecularMIMOBirkan} that the adaptive thresholding mechanisms work better than FTD approaches for RC, in the presence of relatively high ISI. 

\par Fig. \ref{fig:single_Msweep} also implies that SMUX faces a high error floor, even though the scheme can transmit at a symbol duration of $8\times 0.25 = 2$ s and circumvents the ISI introduced by the MCvD channel. The reason for this high error floor of SMUX is the significant ILI. Note that, since every antenna pair transmits independent streams, the random walks of the molecules cause some of them to arrive at other receiver antennas rather than their intended antennas. This imposes heavy ILI for SMUX, creating an irreducibly high error floor. 

\par Overall, it can be seen that $8$-MSSK performs considerably better than the SISO baseline, SMUX, and RC approaches. One reason behind this behavior is the fact that $8$-MSSK is able to transmit less frequently and with more molecules on each transmission while still satisfying the energy consumption and bit rate constraints. Since $8$-MSSK is able to embed three bits in every transmission, it can transmit at a symbol duration of $3t_b$ and with $\frac{3}{2}M^{Tx}$ molecules. Compared to the $t_b$ duration and $M^{Tx}$ molecules of the SISO baseline, it can be inferred that $8$-MSSK faces comparatively less ISI and relative arrival variance \cite{arrivalmodel}, lowering its BER. 

\par The major reason of $8$-MSSK's better performance lies in the fact that it inherently lowers ILI. When a transmission is made from a certain transmitter antenna, the antennas in the corresponding receiver antenna's vicinity also receive a non-negligible number of molecules (also shown in Table \ref{tab:coefs}). Since $8$-MSSK uses only one out of the available eight antennas per transmission, the vicinities of the receiver antennas that are spatially further away from the intended antenna are able to get \textit{cleaned} from the residual molecules, which would otherwise cause ILI. This phenomenon keeps the overall ILI lower at the receiver end, and makes MSSK very suitable for molecular MIMO systems, which experience ISI and ILI otherwise. 

\par Overall, it can be concluded that MSSK provides efficient ISI and ILI combating for a molecular MIMO system and yields a consistent downward slope for BER as $M^{Tx}$ increases. Acknowledging the relation between MSSK's error performance and the ISI/ILI a molecular MIMO system faces, the rest of this subsection analyzes MSSK's BER behavior under varying bit rate constraints and antenna separations.

\subsubsection*{Effect of the Bit Rate Constraint}
\label{subsec:single_tb}

\par
\begin{figure}[h]
	\centering
	\includegraphics[width=0.48\textwidth]{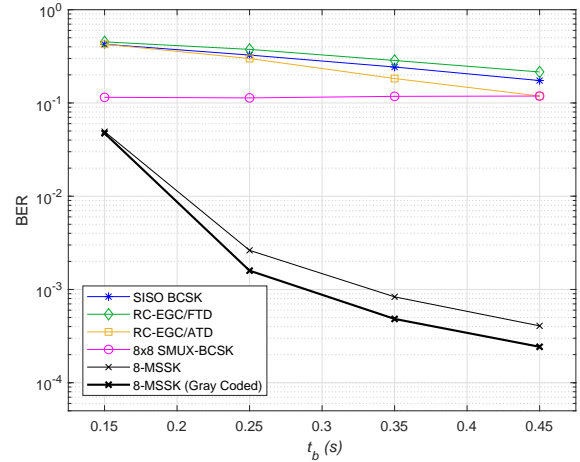} 
	\caption{BER vs. $t_b$ curves for the single-molecule MIMO approaches. $M^{Tx} = 300$, $d_x = 10\mu$m, $d_{yz} = 10\mu$m, $D = 79.4 \frac {\mu m^{2}}{s}$, and $r_r = 5\mu$m. }
	\label{fig:single_tbsweep}
\end{figure}

 The bit rate of any MCvD system directly affects the ISI it experiences at the receiver end \cite{tb_ISI}. In order to analyze the effects of the bit rate constraint of the MIMO approaches analyzed in Fig. \ref{fig:single_Msweep}, Fig. \ref{fig:single_tbsweep} is presented. 

\par Fig. \ref{fig:single_tbsweep} shows that the BER performance of $8$-MSSK faces an error floor, in which increasing $t_b$ yields diminishing returns in terms of lowering BER. The reason for this behavior lies in the presence of ILI. Even though MSSK inherently reduces ILI, ILI still exists to some extent since it is a physical implication of the MCvD channel, causing the error floor in Fig. \ref{fig:single_Msweep}. 

\par Furthermore, the gap between natural binary and Gray mapped $8$-MSSK increases as $t_b$ increases. Recalling that Gray coding is applied solely for reducing bit errors when an ILI-caused symbol error occurs, it may be concluded that ILI even gets slightly worse when $t_b$ is increased. This is due to the fact that when the symbol duration increases (as a result of increasing $t_b$), the arrival distribution of the molecules become more \textit{balanced} among the antennas. Hence, even though waiting too long for the molecules to arrive is beneficial in terms of reducing ISI, the balancing effect creates slightly more ILI at the receiver end as it smooths the distinct largeness of the intended antenna's channel coefficient. It should, however, be noted that the ISI reduction of increasing $t_b$ is much more significant than the slight ILI increase, causing the overall downward trend in BER.

\par The slight ILI increase can also be validated from the fact that SMUX-BCSK's BER slightly increases as $t_b$ increases, unlike other molecular MIMO schemes. Note that since SMUX transmits at a rate of $8t_b$, it faces very little ISI to begin with. This implies that the scheme's errors are mainly caused by ILI. SMUX-BCSK's slightly increasing BER with $t_b$ verifies the slight growth in ILI as $t_b$ increases.

\subsubsection*{Effect of Antenna Separation}

\par
\begin{figure}[h]
	\centering
	\includegraphics[width=0.48\textwidth]{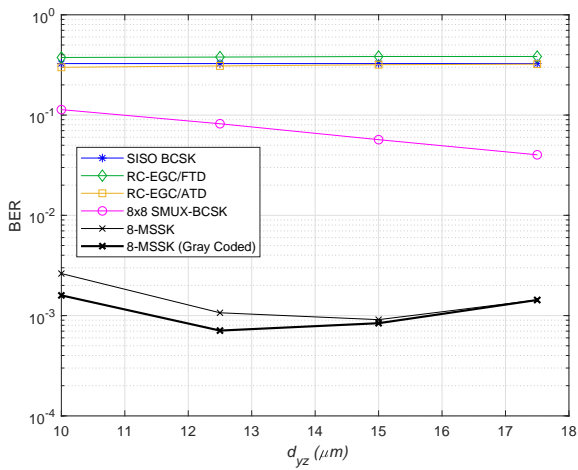} 
	\caption{BER vs. $d_{yz}$ curves for the single-molecule MIMO approaches. $M^{Tx} = 300$, $t_b = 0.25$s, $d_x = 10\mu$m, $D = 79.4 \frac {\mu m^{2}}{s}$, and $r_r = 5\mu$m.}
	\label{fig:single_dyzsweep}
\end{figure}

 As also discussed in \cite{molecularMIMOBirkan}, the ILI faced in a molecular MIMO system is significantly affected by the spatial separation between antennas. Since the system of interest uses a UCA with a distance of $d_{yz}$ from the center for each receiver antenna, the antenna separation is determined by the parameter $d_{yz}$ in this paper. Fig. \ref{fig:single_dyzsweep} presents the effects of antenna separation on BER.

\par Fig. \ref{fig:single_dyzsweep} implies an interesting result: Antenna separation hurts MSSK after a certain point. Even though an increase in $d_{yz}$ reduces ILI significantly, it actually adds some ISI into the system. The argument presented in \cite{multrec_absorb} can be stated to explain this phenomenon: Nearby absorbing receivers actually help reduce the ISI by absorbing the \textit{astray} molecules that generally take longer to arrive at the intended antenna. An increase in $d_{yz}$ reduces this cancellation effect, and worsens the ISI combating of nearby antennas. Until a certain point, the reduction in ILI dominates the increase in ISI caused by increasing $d_{yz}$. However, after the mentioned point, the channel becomes ISI-dominated (with negligible ILI), and increasing $d_{yz}$ further hurts the system. Note that this effect is not significant for SMUX, as it is a very heavily ILI-dominated scheme. Since ISI is very low and ILI is very high for SMUX to begin with, increasing $d_{yz}$ generally helps the approach.

\subsubsection*{Effect of Flow}

\par
\begin{figure}[h]
 	\centering
 	\includegraphics[width=0.48\textwidth]{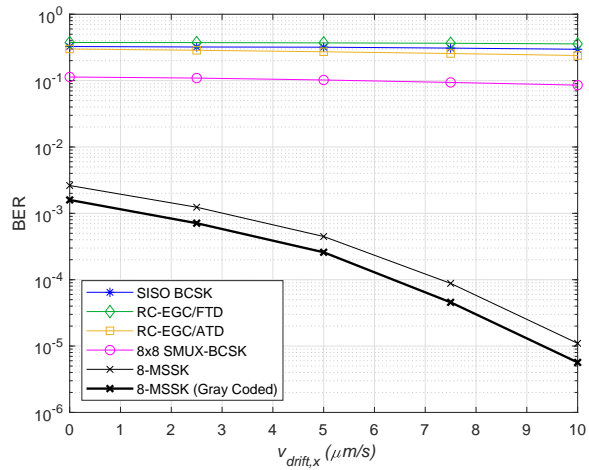} 
 	\caption{BER vs. drift velocity ($v_x$) curves for the single-molecule MIMO approaches. $M^{Tx} = 300$, $t_b = 0.25$s, $d_x = 10\mu$m, $D = 79.4 \frac {\mu m^{2}}{s}$, and $r_r = 5\mu$m.}
	\label{fig:single_drift}
\end{figure}

 As also mentioned in Section \ref{systemmodel}, a 3-D molecular communication environment without drift is considered throughout this paper. However, the error performance of molecular communication systems changes in the presence of flow in the channel \cite{birkansurvey}. Motivated by this, a uniform flow is applied to the system presented in Fig. \ref{fig:UCA} in the positive $x$-axis (with drift velocity $v_{\text{drift,}x}$), and MSSK's error performance is comparatively analyzed with the existing molecular MIMO schemes on Fig. \ref{fig:single_drift}.

\par Overall, the results of Fig. \ref{fig:single_drift} show that increasing the drift velocity towards the receiver benefits the communication performance. The reason behind this trend is that a positive drift towards the receiver causes more molecules to take shorter times to arrive at the receiver. The molecules arriving quicker at the receiver mitigates ISI on the receiver end, which in turn reduces BER. 

\subsection{Dual-Molecule Systems}
\label{subsec:errortwo}

\par
\begin{figure}[h]
	\centering
	\includegraphics[width=0.48\textwidth]{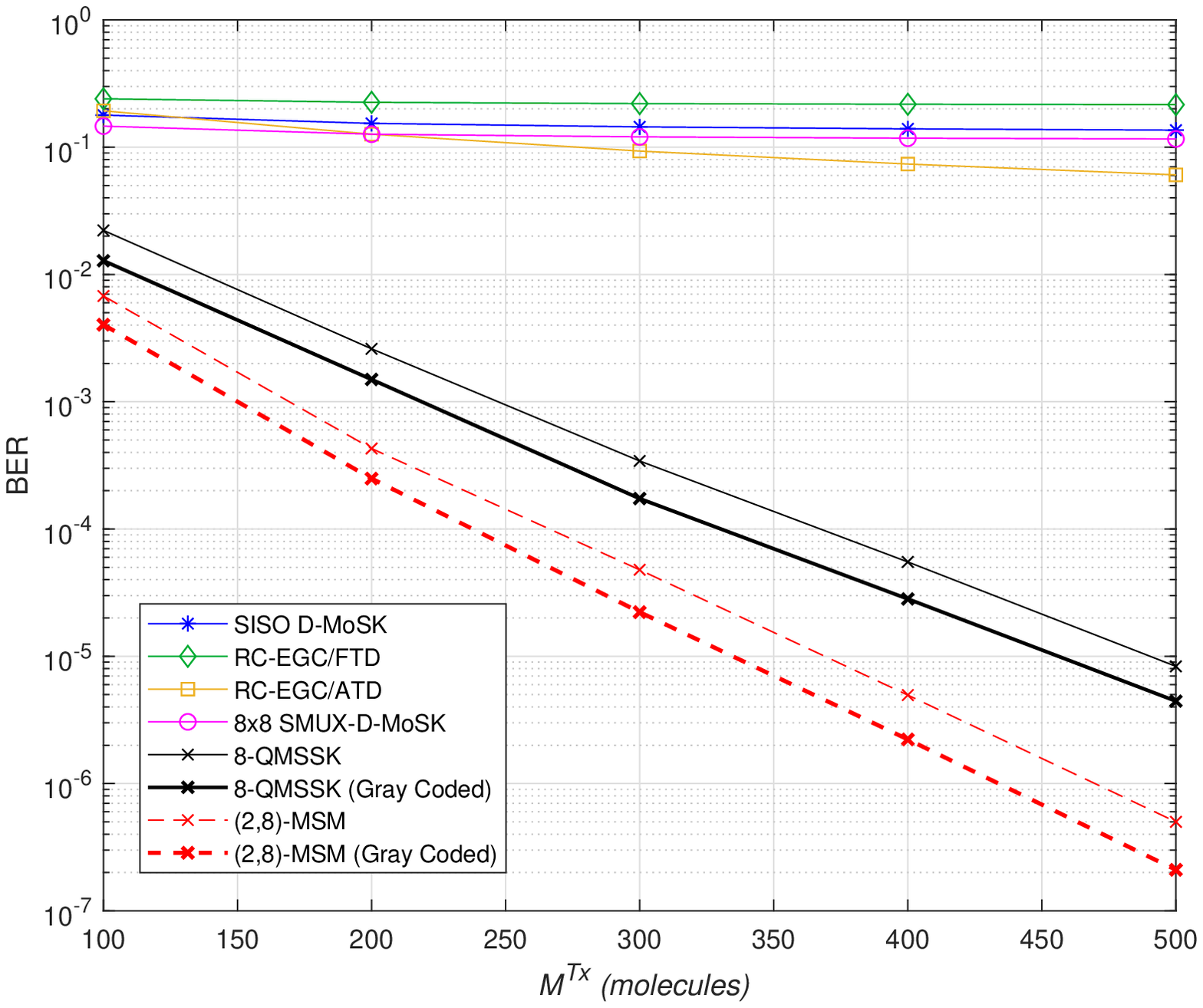} 
	\caption{BER vs. $M^{Tx}$ curves for the dual-molecule MIMO approaches. $t_b =0.25$s, $d_x = 10\mu$m, $d_{yz} = 10\mu$m, $D = 79.4 \frac {\mu m^{2}}{s}$, and $r_r = 5\mu$m. }
	\label{fig:two_Msweep}
\end{figure}

 When a second type of molecule is introduced to the system, the increased degree of freedom can be utilized to enhance the error performance. As discussed in Subsection \ref{subsec:two}, the proposed dual-molecule IM schemes are QMSSK and MSM. Recalling $n_{Tx} = n_{Rx} = 8$ for this paper, the employed schemes are $8$-QMSSK, and ($2,8$)-MSM. At the receiver end, \eqref{eq:MSSKmax} is used as the decoder for $8$-QMSSK, and \eqref{eq:MSMmax} is employed for ($2,8$)-MSM. It can also be recalled from Subsection \ref{subsec:two} that the SISO baseline modulation becomes binary D-MoSK, since it is the direct extension of SISO BCSK to two types of molecules. RC and SMUX are also assumed to create two orthogonal and parallel channels using the two types of molecules. Overall, the BER performance of the aforementioned schemes with respect to $M^{Tx}$ is presented in Fig. \ref{fig:two_Msweep}.

\par Similar to the discussion made for $8$-MSSK in Subsection \ref{subsec:errorsingle}, the IM-based schemes perform considerably better than SISO baseline, SMUX, and RC approaches. Note that $8$-QMSSK is able to transmit at a symbol duration of $6t_b$ and using $\frac{3}{2}M^{Tx}$ molecules per transmission per molecule, and ($2,8$)-MSM is able to transmit with $4t_b$ and using $\frac{4}{2}M^{Tx} = 2M^{Tx}$. Furthermore, since $8$-QMSSK and ($2,8$)-MSM are both antenna index-based modulations, they inherently lower ILI with the cleaning effect as discussed in Subsection \ref{subsec:errorsingle}. Both $8$-QMSSK and ($2,8$)-MSM face less overall interference and arrival noise, and provide less bit errors than other approaches. 

\par Fig. \ref{fig:two_Msweep} also shows that ($2,8$)-MSM achieves a better error performance than $8$-QMSSK. Even though $8$-QMSSK transmits with a symbol duration of $6t_b$ and has less ISI, ($2,8$)-MSM combats ILI much better than $8$-QMSSK without losing significantly from its ISI combating capability while transmitting at $4t_b$. Furthermore, ($2,8$)-MSM is able to transmit more molecules per channel use per molecule type, which in turn lowers its relative arrival variance according to the findings of \cite{arrivalmodel}. All in all, much better ILI combating while not losing significantly from ISI combating makes ($2,8$)-MSM surpass $8$-QMSSK, error performance-wise. 

\subsubsection*{Effect of the Bit Rate Constraint}

\par
\begin{figure}[h]
	\centering
	\includegraphics[width=0.48\textwidth]{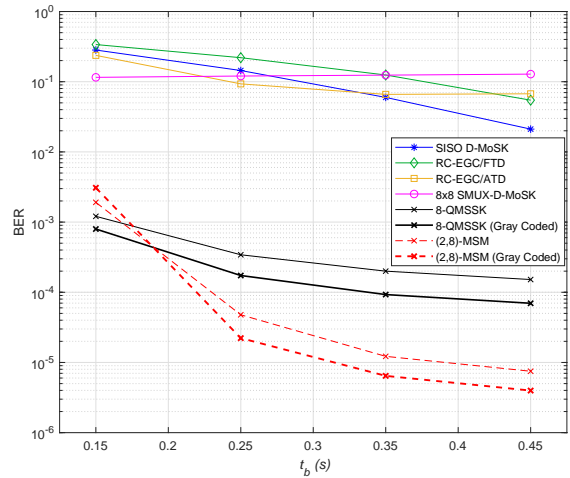} 
	\caption{BER vs. $t_b$ curves for the dual-molecule MIMO approaches. $M^{Tx} = 300$, $d_x = 10\mu$m, $d_{yz} = 10\mu$m, $D = 79.4 \frac {\mu m^{2}}{s}$, and $r_r = 5\mu$m. }
	\label{fig:two_tbsweep}
\end{figure}

 Similar to the single-molecule scenarios, the bit rate constraint is a major factor in the amount of ISI the dual-molecule MIMO schemes face as well. The effects of the bit duration constrainti $t_b$, are presented in Fig. \ref{fig:two_tbsweep}. 

\par From Fig. \ref{fig:two_tbsweep}, it can be seen that ($2,8$)-MSM performs worse than $8$-QMSSK for lower $t_b$ values. This behavior is due to the fact that $8$-QMSSK combats ISI better since it transmits with a symbol duration of $6t_b$, compared to ($2,8$)-MSM's $4t_b$ duration. Note that ($2,8$)-MSM's errors at $t_b = 0.15$s are ISI-caused, as it is also validated by the fact that Gray coding fails to reduce BER compared to the binary code. At higher data rates, better ISI combating allows $8$-QMSSK to maintain its reliability better than ($2,8$)-MSM. However, as $t_b$ increases, the higher ILI imposed on $8$-QMSSK causes a higher error floor than ($2,8$)-MSM's. 

\section{Receiver Design}
\label{sec:receiverdesign}

\subsection{Maximum Count Decoder}
\par As \eqref{eq:MSSKmax} suggests, performing maximum count decoding on the antenna arrival vector is a computationally efficient method of decoding, since MCD is memoryless and it does not require access to the CIR matrix presented in Table \ref{tab:coefs}. It is noteworthy that even with  such a simple detector, it is demonstrated in Section \ref{sec:errorperformance} that MSSK still outperforms the existing molecular MIMO schemes by yielding a steeper slope with respect to $M^{Tx}$ in Fig. \ref{fig:single_Msweep} and lower BER values overall. However, given that nano-machines are equipped with enough computational power and access to CIR using a method like in \cite{estimationMolComMIMO}, better detectors that yield even lower error rates can be constructed at the price of computational complexity. 

\subsection{Maximum Likelihood Sequence Detector}

\par Similar to the maximum likelihood (ML) sequence detection algorithm proposed to the molecular communications literature by \cite{molcom_receivers_akan}, an ML-based sequence detector is also an option in molecular IM schemes, as the MCvD channel has signal-dependent characteristics \cite{sig-dependent-noise}. Assuming perfect CIR at the receiver end, the detector can be thought of as the direct extension of the ML sequence detector presented in \cite{molcom_receivers_akan}, for a symbol alphabet with cardinality $n_{Rx}$ instead of two. The decision rule for such a detector can be expressed as
\begin{equation}
\hat{x}[k-L+1:k] = \argmaxA_{\forall x[k-L+1:k]} \mathcal{L}\Big(\textbf{Q} \big| x[k-L+1:k] \Big)
\label{eq:MLsequence}
\end{equation}
where $x[k-L+1:k]$ defines the activated antenna index vector as mentioned in Section \ref{sec:theo}, and \textbf{Q} denotes the $n_{Rx}$-by-$L$ antenna arrival count matrix for each receiver antenna corresponding to the $x[k-L+1:k]$ index sequence. In \eqref{eq:MLsequence}, $\mathcal{L}\Big(\textbf{Q} \big| x[k-L+1:k] \Big)$ denotes the likelihood function corresponding to a particular symbol sequence $x[k-L+1:k]$. Note that the considered symbol alphabet with cardinality $n_{Rx}$ represents the antenna indices for $n_{Tx}$-MSSK, given $n_{Tx} = n_{Rx}$. 

\par Recalling from Subsection \ref{subsec:channelcoeff}, the Gaussian approximation of the Binomial arrival distribution is valid for the scenarios considered in this paper. Considering Gaussian arrivals, \cite{molcom_receivers_akan} proposes the branch metric for the ML sequence decoder's trellis as
\begin{equation}\label{eq:branchmetricMLseq}
\mathcal{M} (R_j[z],x[k-L+1:k]) = \ln \sigma^2_j[z] + \frac{R_j[z]-\mu_j[z]}{\sigma^2_j[z]}.
\end{equation}
In this expression, $R_j[z]$ represents the received number of molecules for the $j^{th}$ antenna at the $z^{th}$ symbol interval. $\mu_j[z]$ and $\sigma^2_j[z]$ denote the theoretical mean and the variance of the Gaussian distribution associated with $R_j[z]$, given a particular $x[k-L+1:k]$ path in the trellis. Note that the computation of $\mu_j[z]$ and $\sigma^2_j[z]$ requires information about the CIR. Furthermore, the sub-optimal squared Euclidean distance branch metric employed in \cite{molecularMIMOBirkan} and other approaches may also be utilized to calculate the branch metric $\mathcal{M} (R_j[z],x[k-L+1:k])$. With the branch metrics as shown in \eqref{eq:branchmetricMLseq}, \eqref{eq:MLsequence} can be equivalently written as
\begin{equation}\label{eq:MLseq_alt}
\begin{split}
\hat{x}[k&-L+1:k] =\\ &\argminA_{\forall x[k-L+1:k]}  \sum_{z=k-L+1}^{k} \sum_{j=1}^{n_{Rx}} \mathcal{M}(R_j[z],x[k-L+1:k]).
\end{split}
\end{equation}

\par The ML sequence detector is equivalent to the maximum aposteriori probability (MAP) sequence detector when the transmission probabilities all symbols in the alphabet are equal, which is the case in this paper as the probability of occurrence of a bit-1 is considered $\frac{1}{2}$ \cite{proakis_kitap}. However, the ML sequence detector needs to generate the trellis and find the likelihood of all $n_{Rx}^{\hspace{0.1cm}L}$ possible $x[k-L+1:k]$ combinations, making the complexity of the scheme proportional to $\mathcal{O}\big((n_{Rx})^{L}\big)$ for its use on MCvD index modulations. Note that even though the Viterbi algorithm with a smaller memory $L_{v}$ than $L$ may be utilized to reduce computational complexity at the cost of losing detection accuracy, the algorithm is still computationally intensive for a nano-machine to handle $\mathcal{O}\big((n_{Rx})^{L_v}\big)$. As also mentioned in \cite{molcom_receivers_akan}, this algorithm's requirements drastically increases the computational complexity, even for binary communications. Recall that the channel memory $L$ is chosen as $L=30$ for accurate representation of the channel and the number of antennas is $n_{Rx}=8$ for the paper. With this in mind, performing $8^{30} \approx 1.2 \times 10^{27}$ operations for detecting a single symbol sequence is a substantial and an almost impossible computational burden for a nano-scale machine. Considering the nano-machines are small devices with limited computational capacity, the computational impracticality of the ML sequence detector hinders its possible use in this paper's scenarios of interest.


\subsection{Symbol-by-Symbol Maximum Likelihood Detector} 

\par Given the impractically high computational complexity of the ML sequence detector, a decoder that works in a symbol-by-symbol manner is beneficial for nano-scale machinery. Combining this idea and the ML concept for decoders, this subsection theorizes a symbol-by-symbol ML detector for the $n_{Tx}$-MSSK modulation considering the availability of CIR at the receiver end. The scheme is referred to as the Symbol-ML detector throughout the paper. 

\par For a certain channel memory $L$, the Symbol-ML detector holds the last $L-1$ decisions as $\hat{x}[k-L+1:k-1]$, and generates the estimated arrival mean and variance for each antenna depending on the past decisions. The estimated total arrival mean and variance on the $j^{th}$ receiver antenna that is caused by the past transmissions can be expressed as
\begin{equation}\label{eq:pastmean_ML}
\hat{\mu}_{j,\text{past}}[k] = \sum_{z=k-L+1}^{k-1} \sum_{i=1}^{n_{Tx}} \hat{s}_{i}[z] h_{i,j}[k-z+1]
\end{equation}
and 
\begin{equation}\label{eq:pastvar_ML}
\begin{split}
\hat{\sigma}^2&_{j,\text{past}}[k] = \\ &\sum_{z=k-L+1}^{k-1} \sum_{i=1}^{n_{Tx}} \hat{s}_{i}[z] h_{i,j}[k-z+1] \big(1-h_{i,j}[k-z+1]\big),
\end{split}
\end{equation}
in a manner similar to \eqref{eq:gausmean} and \eqref{eq:gausvar}. Recalling that $x[k]$ denotes the activated antenna for the $k^{th}$ transmission instant for MSSK, $\hat{s}_{i}[k] = \frac{\log_2n_{Tx}}{2}M^{Tx}$ if $\hat{x}[k] = i$, and is zero otherwise.

\par After determining $\hat{\mu}_{j,\text{past}}[k]$ and $\hat{\sigma}^2_{j,\text{past}}[k]$, the detector calculates the estimated mean and variance vectors given $\hat{x}[k] = i$ is true. Denoting these vectors as $\boldsymbol{\hat{\mu}}_ {\textbf{i}} \textbf{[k]}$ and $\boldsymbol{ {\hat{\sigma}}}^2_{\textbf{i}} \textbf{[k]}$, respectively, 
\begin{equation}\label{eq:mean_ML}
\boldsymbol{\hat{\mu}}_ {\textbf{i}} \textbf{[k]} = 
\begin{bmatrix}
\hat{\mu}_{1,\text{past}}[k] + s_{\text{MSSK}} h_{i,1}[1] \\
\hat{\mu}_{2,\text{past}}[k] + s_{\text{MSSK}} h_{i,2}[1] \\
\vdots \\
\hat{\mu}_{n_{Rx},\text{past}}[k] + s_{\text{MSSK}} h_{i,n_{Rx}}[1] \\
\end{bmatrix}
\end{equation}
and
\begin{equation}\label{eq:var_ML}
\boldsymbol{\hat{\sigma}}^2_{\textbf{i}} \textbf{[k]} =
\begin{bmatrix}
\hat{\sigma}^2_{1,\text{past}}[k] + s_{\text{MSSK}} h_{i,1}[1] (1-h_{i,1}[1]) \\
\hat{\sigma}^2_{2,\text{past}}[k] + s_{\text{MSSK}} h_{i,2}[1] (1-h_{i,2}[1])\\
\vdots \\
\hat{\sigma}^2_{n_{Rx},\text{past}}[k] + s_{\text{MSSK}} h_{i,n_{Rx}}[1](1-h_{i,n_{Rx}}[1]) \\
\end{bmatrix}
\end{equation}
are computed. Note that this operation is performed for $i = 1,...,n_{Rx}$, generating $n_{Tx}$ number of $n_{Rx}$-by-1 vectors. Furthermore, recall that $s_{\text{MSSK}} = \frac{\log_2n_{Tx}}{2}M^{Tx}$. Denoting the $j^{th}$ element of $\boldsymbol{\hat{\mu}}_ {\textbf{i}} \textbf{[k]}$ as $(\boldsymbol{\hat{\mu}}_ {\textbf{i}} \textbf{[k]})_j$, the log-likelihood function is applied on each receiver antenna for each $\hat{x}[k]=i$, to yield 

\begin{equation}\label{loglik_ML}
(\boldsymbol{\mathcal{H}^i})_{j} =   \ln \Bigg( \frac{1}{\sqrt{2\pi \boldsymbol{\hat{\sigma}}^2_{\textbf{i}} \textbf{[k]}_j}} \Bigg)  - \frac{\Big(R_j[k] - (\boldsymbol{\hat{\mu}}_ {\textbf{i}} \textbf{[k]})_j\Big)^2}{2(\boldsymbol{\hat{\sigma}}^2_{\textbf{i}} \textbf{[k]})_j} 
\end{equation}
where $\boldsymbol{\mathcal{H}^i}$ represents the log-likelihood vector given $\hat{x}[k]=i$ is true. Lastly, the decoded symbol $\hat{x}[k]$ is found by finding the maximum sum among all $\boldsymbol{\mathcal{H}^i}$ by performing
\begin{equation}\label{l1_norm_ML}
\hat{x}[k] = \argmaxA_{i \in 1,...,n_{Rx}} \sum_{j=1}^{n_{Rx}}(\boldsymbol{\mathcal{H}^i})_j.
\end{equation}

\par Note that similar to the MCD, this detector can also be directly extended to $n_{Tx}$-QMSSK and ($\beta,n_{Tx}$)-MSM. The procedure can be done separately on the two types of molecules when decoding QMSSK, and the detector may be extended to an alphabet with $\log_2 (\beta) + \log_2 (n_{Tx})$ elements for the detector of MSM.

\par
\begin{figure}[h]
	\centering
	\includegraphics[width=0.48\textwidth]{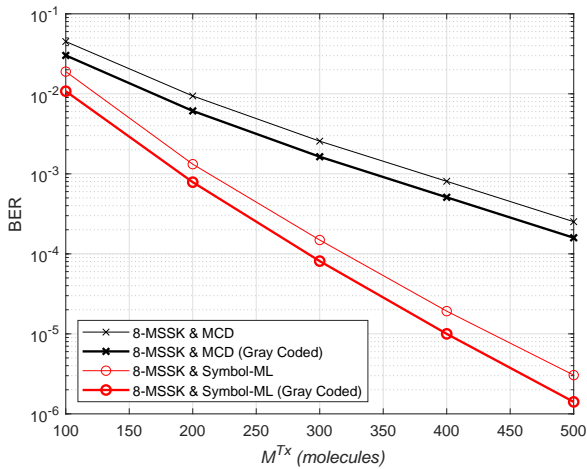} 
	\caption{BER vs. $M^{Tx}$ curves natural binary and Gray mapped $8$-MSSK using the MCD and Symbol-ML decoders. $t_b = 0.25$s, $d_x = 10\mu$m, $d_{yz} = 10\mu$m, $D = 79.4 \frac {\mu m^{2}}{s}$, and $r_r = 5\mu$m. }
	\label{fig:ML_BER}
\end{figure}

 Overall, symbol-by-symbol ML decoder requires access to the CIR, requires more computation power than MCD even though it is much less complex than ML sequence decoder, and may have error propagations under bad channel conditions due to its decision feedback nature. The Symbol-ML's and MCD's error performances for $8$-MSSK are comparatively analyzed in Fig. \ref{fig:ML_BER}. Note that the same channel parameters as in Fig. \ref{fig:single_Msweep} are used, and $8$-MSSK is employed as the molecular IM.

\par Fig. \ref{fig:ML_BER} shows that Symbol-ML yields lower error rates with steeper slopes than MCD, despite its potential error propagation problem. The better error performance of Symbol-ML is mainly due to its access to CIR. Compared to the crude maximization MCD does on $R_j[k]$'s, the added complexity and the channel information allow Symbol-ML to perform a more elaborate and channel-aware decoding and helps it to reduce BER.

\section{Conclusion}
\label{sec:conclusion}

\par In this study, IM-based concepts have been introduced to the field of molecular communications. Molecular IM schemes that are suitable for single and multiple available molecule types have been proposed in the paper, and it has been found that said modulations yield very promising results for molecular MIMO systems. Proposed IM schemes overcome the ISI problem better than the available space-time coding approaches, mainly due to their ability to encode multiple bits at a single transmission. Furthermore, encoding bits in the antenna index has been observed to combat ILI very effectively, which is generally a limiting factor for SMUX-based systems. Overall, the proposed modulations have been found to yield low error probabilities while conserving high data rates for MCvD-MIMO systems. 

\par Due to the MCvD channel's physical nature, a trade-off between ISI and ILI has been pointed out in the paper, and it has been observed that the error performance does not always improve by increasing antenna separation for IM schemes. Gray coding for antenna indices has been found to decrease error rates for the vast majority of the cases, and has been acknowledged to be a very useful addition for molecular IM schemes to combat ILI-caused errors. Furthermore, as expected, introducing the second molecule type to a molecular MIMO system has been observed to be useful for molecular MIMO systems, as it allows the transmitter nano-machine to perform more elaborate index modulations such as QMSSK and MSM, further lowering the error probability. 

\par Lastly, since this paper's main goal is to introduce the IM concept to the field of molecular communications by proposing single and dual-molecule IM schemes, possible issues regarding misalignments between antennas, temporal variations, and other imperfections are outside the scope of this paper. Alongside the development of other molecular IM-based schemes and receiver designs, characterization and counter-measures regarding these possible issues are left as future works.

\section*{Appendix}

\par Here, the particle-based simulation algorithm used to evaluate the bit error rate of the MSSK scheme is presented. Note that only the $L^{th}$ symbol is taken into account on the error probability calculation for a channel of memory $L$. This approach is employed in order to avoid an overly-optimistic result by ensuring each symbol considered in the evaluation is subjected to the full channel memory (hence ISI) of $L$.

\begin{algorithm}[]
		\fontsize{10}{10}\selectfont
		\begin{algorithmic}[1]
			\caption{Algorithm for the particle-based simulation to evaluate the bit error rate of the MSSK scheme.}
			\label{alg:particle}
			\renewcommand{\algorithmicrequire}{\textbf{Inputs:}}
			\renewcommand{\algorithmicensure}{\textbf{Output:}}
			\REQUIRE $L$, $t_b$, $\Delta t$, $D$, $n_{Tx}$, $n_{Rx}$, $r_r$, $d_x$, $d_{yz}$ \\
			$n_{\text{trials}}$: Total number of trials for Monte Carlo analysis\\
			$\frac{M^{Tx}}{2}$: Molecule budget to transmit a single bit \\
			$\textbf{R}$: Received molecule count matrix for each RX antenna and symbol interval\\
			$[\textbf{x}_{Tx},\textbf{y}_{Tx},\textbf{z}_{Tx}]$,  $[\textbf{x}_{Rx},\textbf{y}_{Rx},\textbf{z}_{Rx}]$: Coordinates of the TX and RX antennas, respectively \\
			
			\ENSURE  Bit error rate ($P_e$)
			\\ \textit{Initialization}: Number of bit errors $N_e = 0$ \\
			\STATE Symbol duration $t_s = \log_2{(n_{Tx})}t_b$\\
			\FOR {$u = 1$ to $n_{\text{trials}}$}
			\STATE Randomly generate $L$ symbols
			\FOR {$m = 1$ to $L t_s \Delta t $}
			\IF {Beginning of the $k^{th}$ symbol interval}
			\STATE Emit $\frac{\log_2{(n_{Tx})} M^{Tx}}{2}$ molecules from the activated antenna's coordinates
			\ENDIF
			\FOR {$i = 1$ to $L \frac{\log_2{(n_{Tx})} M^{Tx}}{2}$ }
			\IF {$i^{th}$ molecule is emitted \& not yet absorbed}
			\STATE Compute $\Delta X, \Delta Y, \Delta Z \sim \mathcal{N}(0,2D\Delta t)$
			\STATE Update position: $x_i \leftarrow x_i + \Delta X$; \hspace{0.05cm} $y_i \leftarrow y_i + \Delta Y$; \hspace{0.05cm} $z_i \leftarrow z_i + \Delta Z$
			\algstore{myalg}
	\end{algorithmic}
\end{algorithm}
\begin{algorithm}[]
	\fontsize{10}{10}\selectfont
	\begin{algorithmic}[1]
			\algrestore{myalg}
			\IF {$i^{th}$ molecule crossed a reflective surface boundary}
			\STATE Perform elastic collision to correct current position
			\ENDIF
			\FOR {$j=1$ to $n_{Rx}$}
			\IF {$\norm{(x_i,y_i,z_i) - (\textbf{x}_{Rx,j},\textbf{y}_{Rx,j},\textbf{z}_{Rx,j})} < r_r$ }
			\STATE $\textbf{R}_j[k] \leftarrow \textbf{R}_j[k] + 1$ (absorption)
			\STATE Flag the $i^{th}$ molecule as absorbed
			\ENDIF
			\ENDFOR		
			\ENDIF 
			\ENDFOR 
			
			\ENDFOR 	
			\STATE $\hat{x}[L] = \arg \max(\textbf{R}_1[L],\dots,\textbf{R}_{n_{Rx}}[L])$ to decode the $L^{th}$ symbol 
			\STATE Map $\hat{x}[L]$ to the $\log_2{n_{Rx}}$-bit long bit sequence (different for natural and Gray mapping)
			\STATE Compute the number of bit errors $e_u$ by comparing with the original bit sequence (different for natural and Gray mapping)
			\STATE $N_e \leftarrow N_e + e_u$
			\ENDFOR \\
			\textbf{return} $P_e = \frac{N_e}{n_{\text{trials}} \log_2(n_{Tx})}$
		\end{algorithmic}
\end{algorithm}

\newpage



\begin{IEEEbiography}[{\includegraphics[width=1in,height=1.25in,clip,keepaspectratio]{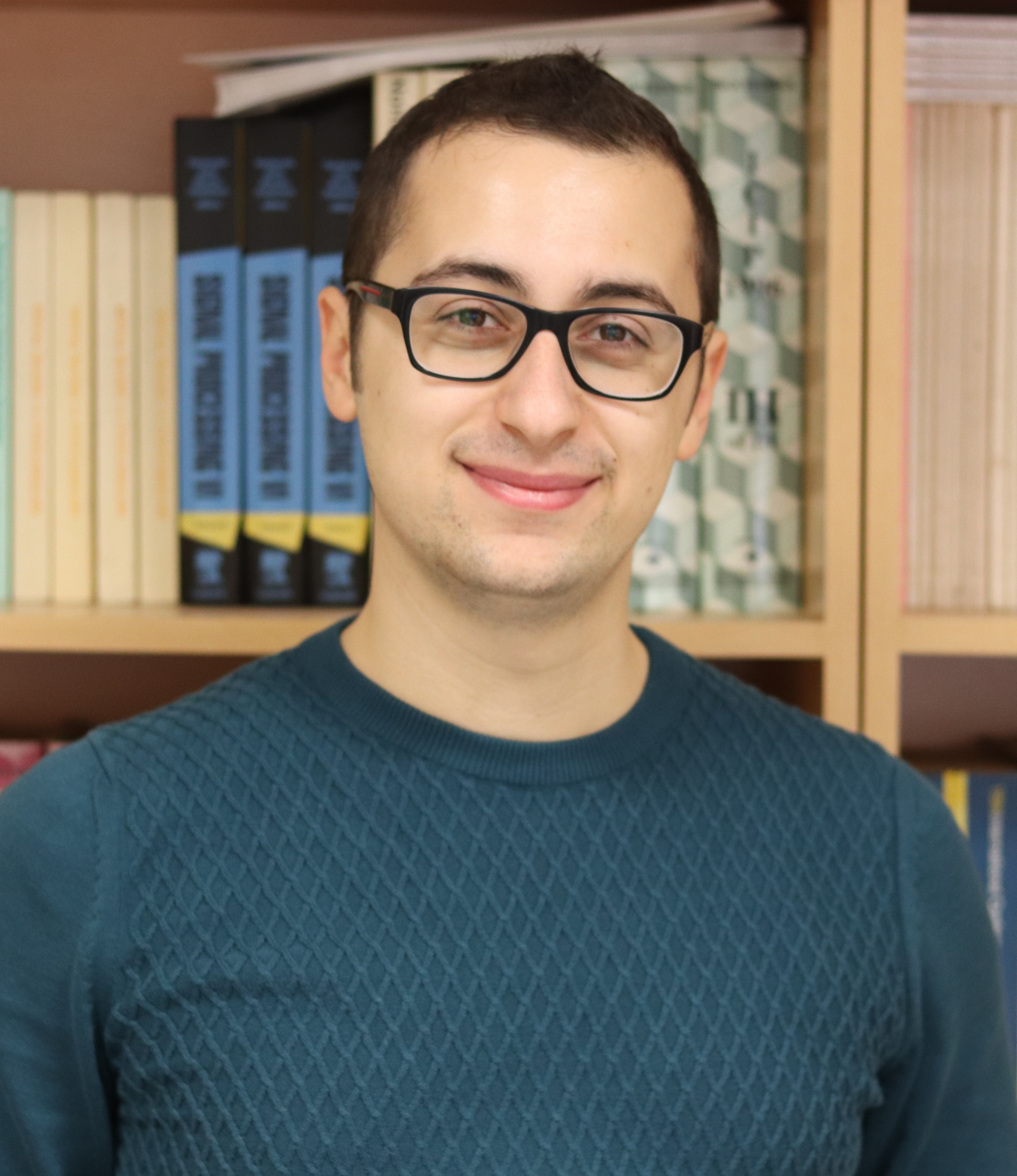}}]{Mustafa Can Gursoy} received the B.Sc. and M.Sc. degrees in electrical and electronics engineering from Bogazici University, Istanbul, Turkey, in 2015 and 2017, respectively. He is currently a Ph.D. student and a research assistant working at the Department of Electrical and Electronics Engineering, Bogazici University, Istanbul, Turkey. His research interests include channel characteristics, modulations, channel coding, and networking approaches for molecular communications.
\end{IEEEbiography}

\begin{IEEEbiography}[{\includegraphics[width=1in,height=1.25in,clip,keepaspectratio]{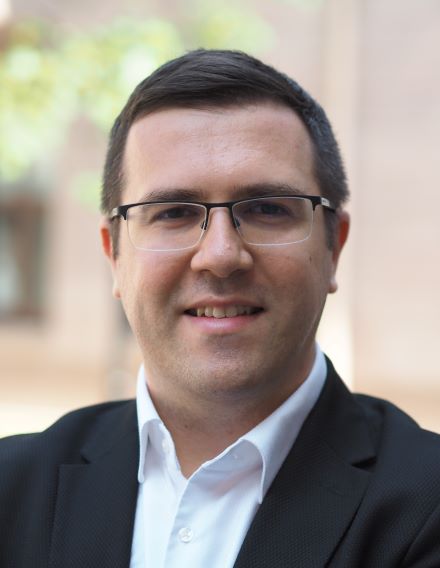}}]{Ertugrul Basar} (S'09-M'13-SM'16) received the B.S. degree (Hons.) from Istanbul University, Turkey, in 2007, and the M.S. and Ph.D. degrees from Istanbul Technical University, Turkey, in 2009 and 2013, respectively. He is currently an Associate Professor with the Department of Electrical and Electronics Engineering, Ko\c{c} University, Istanbul, Turkey and the director of Communications Research and Innovation Laboratory (CoreLab). His primary research interests include MIMO systems, index modulation, waveform design, visible light communications, and signal processing for communications.
\par Dr. Basar currently serves as an Editor of the \textsc{IEEE Transactions on Communications} and \textit{Physical Communication} (Elsevier), and as an Associate Editor of the \textsc{IEEE Communications Letters}. He served as an Associate Editor for the \textsc{IEEE Access} from 2016 to 2018. 
\end{IEEEbiography}
\balance
\begin{IEEEbiography}[{\includegraphics[width=1in,height=1.25in,clip,keepaspectratio]{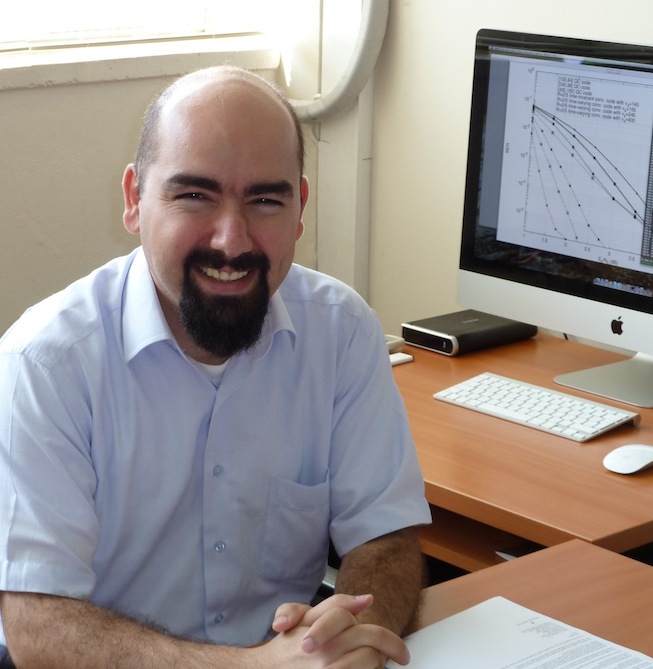}}]{Ali Emre Pusane} received the B.Sc. and M.Sc. degrees in electronics and communications engineering from Istanbul Technical University, Istanbul, Turkey, in 1999 and 2002, respectively, and the M.Sc. degree in electrical engineering, the M.Sc. degree in applied mathematics, and the Ph.D. degree in electrical engineering from the University of Notre Dame, Notre Dame, IN, in 2004, 2006, and 2008, respectively. He was a Visiting Assistant Professor at the Department of Electrical Engineering, University of Notre Dame, during 2008-2009, after which he joined the Department of Electrical and Electronics Engineering, Bogazici University, Istanbul, Turkey. His research is in wireless communications, information theory, and coding theory.
\end{IEEEbiography}

\begin{IEEEbiography}[{\includegraphics[width=1in,height=1.25in,clip,keepaspectratio]{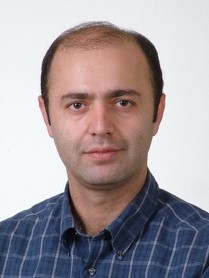}}]{Tuna Tugcu} received the B.S. and Ph.D. degrees in computer engineering from Bogazici University, Istanbul, Turkey, in 1993 and 2001, respectively, and the M.S. degree in computer and information science from the New Jersey Institute of Technology, Newark, NJ, USA, in 1994. He was previously a postdoctoral fellow and a visiting professor with Georgia Institute of Technology, USA. He is currently a professor in the Department of Computer Engineering, Bogazici University. His research interests include nanonetworking, molecular communications, wireless networks, and IoT. Prof. Tugcu has served with the North Atlantic Treaty Organization science and technology groups and the IEEE standards groups.
\end{IEEEbiography}

\end{document}